\documentclass[preprint,10pt,1p,sort&compress]{elsarticle}

\usepackage{amssymb,amsmath}
\usepackage{graphicx}
\usepackage{booktabs}
\usepackage{tabularx}    
\usepackage[font=small,labelfont=bf, justification=justified, format=plain]{caption}
                     
\def\bea{\begin{eqnarray}}
\def\eea{\end{eqnarray}}
\def\be{\begin{equation}}
\def\ee{\end{equation}}
\def\fr{\frac}
\def\lm{\lambda}

\def\la{\label}
\def\be{\begin{equation}}
\def\ee{\end{equation}}
\def\le{\left}
\def\ri{\right}
\journal{Physics of the Dark Universe}

\begin{document}

\begin{frontmatter}

\title{Analytic Fluid Approximation for Warm Dark Matter}                      

\author[MCTP,CONACYT]{Jorge Mastache}
\ead{jhmastache@mctp.mx}

\author[IFUNAM, SPAIN]{Axel de la Macorra}
\ead{macorra@fisica.unam.mx}

\address[MCTP]{Mesoamerican Centre for Theoretical Physics, Universidad Aut\'{o}noma de Chiapas,  Carretera Zapata Km. 4, Real del Bosque, 29040, Tuxtla Guti\'{e}rrez, Chiapas, M\'{e}xico.}
\address[CONACYT]{Consejo Nacional de Ciencia y Tecnolog\'ia, Av. Insurgentes Sur 1582, Colonia Cr\'edito
Constructor, Del. Benito Ju\'arez, 03940, Ciudad de M\'exico, M\'exico.}
\address[IFUNAM]{Instituto de F\'isica, Universidad Nacional Autonoma de M\'exico, Circuito de la Investigaci\'on 
Cient\'ifica Ciudad Universitaria, 04510, CDMX, M\'exico.}
\address[SPAIN]{Instituto de Ciencias del Cosmos, University of Barcelona, ICCUB, Barcelona 08028, Spain.}

\begin{abstract}
\noindent We present the full evolution of the velocity of a massive particle, along with the equation of state we can compute the energy density and pressure evolution for the background evolution. It is also natural to compute the perturbation equations for any massive decoupled particle, i.e. warm dark matter (WDM) or neutrinos, in the fluid approximation. Using this approach we analytically compute the time when the WDM stop being relativistic, $a_{nr}$, which is 2.6\% different respect to the exact Boltzmann solution. Using the fluid approximation the matter power spectrum is computed faster and with great accuracy, the cut-off in structure formation due to the free-streaming ($\lambda_{fs}$) of the particle, characteristic for a WDM particle, is replicated in both matter power spectrum and halo mass function. With this approach, we have a deeper understanding of the WDM physics that lead us to show that the temperature the dark matter can be computed as a function of known properties of the WDM particle. This formulation can be integrated into comprehensive numerical modeling reasonable increasing the performance in the calculations, therefore, we analyze the parameter $a_{nr}$ in a $\Lambda$WDM model using CMB Planck data combined with matter power spectrum data set of WiggleZ, obtaining a lower bound for the WDM mass $m_{\rm wdm} = 70.3$ eV at 86\% confidence, this value is consistent with WiggleZ data set but more data at small scales or a combination with other observations are needed to stronger constrain the mass value of the WDM particle. 
\end{abstract}

\begin{keyword}
Warm Dark Matter \sep Cosmological parameters \sep Dark matter \sep MCMC \sep CMB Constrainst
\end{keyword}

\end{frontmatter}

\section{Introduction}\label{sec:introduction}

The most successful model to describe the Universe is the $\Lambda$CDM model, which supported by observational evidence such as the Cosmic Microwave Background (CMB) anisotropies \cite{Aghanim:2018eyx}, galaxy redshift surveys \cite{Abbott:2016ktf}, type Ia SuperNovae \cite{Betoule:2014frx} reach to the conclusion that the content of the Universe is composed of 65\% dark energy driving the accelerated expansion of the Universe, 31\% dark matter (DM) whose clustering feature influence the large scale structure formation and the rest 4\% is baryonic matter. Therefore is important to have useful tools to study the main components of the Universe in order to get a glance at its nature.

Most DM particles that have been proposed have non-negligible velocities in the early Universe where it is assumed that DM particles are in thermal contact with the primordial bath, decouple when still relativistic, and then become non-relativistic when the Universe is still radiation-dominated when the primordial clumps begin to cluster to form large scale structures. Therefore, the time when DM become non-relativistic is important and directly proportional to its mass. Depending on how large is the mass the DM is known to be cold, CDM, if $m_{\rm cdm} \sim \mathcal{O}$(MeV). This kind of DM particles stop being relativistic and start clustering object at a very early time. DM with a mass around $m_{\rm wdm} \sim \mathcal{O}$(KeV) are known to be warm, WDM, whose main attribute is that its dispersion velocity wipes out some density concentrations of matter and, therefore, induce a cut-off scale into the mass halo function \cite{Archidiacono:2013dua}.

Having a cut-off scale is an appealing dark matter feature because it conciliates observations with theoretical predictions, for instance, the number of satellite galaxies in our Galaxy is smaller than the expected from CDM simulations, the so-called missing satellite problem \cite{Kazantzidis:2003hb, BoylanKolchin:2011de, Klypin:1999uc}. Solutions to this problem had also been pursued through baryon physics - star formation and halo evolution in the galaxy may be suppressed due to some baryonic process and the discussion is still in progress \cite{Garrison-Kimmel:2017zes, Sawala:2015cdf, Pawlowski:2015qta}.
 
Perhaps the two most important quantities to study the cosmological impact of DM are the amount of energy density today $\Omega_{\rm dmo}$, and the time when these particles become non-relativistic (given by the scale factor $a_{\rm nr}$). There may be a third parameter, the velocity dispersion of the DM particles at $a_{\rm nr}$, its value may reflect the nature of the DM, for instance, an abrupt transition to the non-relativistic DM epoch that may suggest of DM subject to a phase transition \cite{Mastache:2019bxu}.

Rough approximation are usually made to account for the evolution when being relativistic ($a<a_{nr}$) and when DM became non-relativistic ($a>a_{nr}$). But here we present a simple analytic approach for a massive particles which is valid for all times characterized by having a non-negligible thermodynamic velocity dispersion \cite{Bode:2000gq, Colin:2000dn, Hansen:2001zv, Viel:2005qj, Dodelson:1993je, Dolgov:2000ew, Asaka:2006nq, Shi:1998km, Abazajian:2001nj, Kusenko:2006rh, Petraki:2007gq, Merle:2015oja, Konig:2016dzg}. With this approach we can compute the fluid approximation for the perturbation equations for any massive particles, i.e. WDM,  given the analytic solution for the energy density evolution we were able to reproduce the most appealing feature of WDM, the cut-off in the matter power spectrum \cite{Bode:2000gq, Colin:2000dn, Hansen:2001zv,Viel:2005qj}.  For example,  a  $m_{\rm wdm} = 3$ keV is analytically computed to have a $a_{nr}=  2.83 \times 10^{-8}$ while the Boltzmann exact solution gives $a_{nr}= 3.18 \times 10^{-8}$, just a 10.8\% difference. In general, the percentage difference between the numerical value obtained from Boltzmann equations and the analytic one of $a_{nr}$ is on average 2.6\% in a mass range 1-10 keV. 

In this work, we make use of natural units, $c = 1$. We present the work as follows: in Sec.\ref{sec:framework} we present the theoretical warm dark matter framework, in which we include the time when the particle became non-relativistic, $a_{nr}$, the WDM temperature and its free-streaming scale. compute the perturbations of the WDM model in Sec. \ref{ssec:pert}, and the mass halo function in Sec.\ref{ssec:preschechter}. We present our conclusions in Sec. \ref{sec:conclusion}.
 
%%%%%%%%%%%%%%%%%%%%%%%%
%%%        SECTION FRAMEWORK        %%%
%\section{BDM Framework}\label{sec:framework}
\section{DM Framework}\label{sec:framework}
%%%%%%%%%%%%%%%%%%%%%%%%

Relativistic particles with peculiar velocity, $v$, and mass $m$ has a momentum ${p}=\gamma mv$, and energy $E^2=p^2+m^2$, where $\gamma\equiv 1/\sqrt{1-v^2}$. Solving for $v$ we obtain,
\begin{equation} \label{eq:vel_gral_eq}
 v  = \frac{p^2}{\sqrt{m^2 + p^2}}
\end{equation}
The particle is relativistic when the velocity is $v \sim 1$ or equivalently when $p \gg m$. The particle is  non-relativistic when $v \ll 1$, this is $p\ll m$. It is common to establish that a particle becomes non relativistic when $p^2=m^2$ \cite{Bode:2000gq}, when this happens, from Eq.\eqref{eq:vel_gral_eq}, the velocity is simply  $ v_{nr}=1/\sqrt{2}$ with $\gamma_{nr}=\sqrt{2} $, thus  $\gamma_{nr} v_{nr}=1$, all quantities with subindex {\it nr} are evaluated at $a_{nr}$. 
 
In an expanding FRW Universe, the momentum of a relativistic particle redshift as  $p(a)= p_{nr}  (a_{nr}/a) = m (a_{nr}/a)$. Therefore, the velocity at all times in an expanding Universe evolves as
\begin{equation}\label{eq:veldm}
  v(a)  = \frac{ (a_{nr}/a)}{\sqrt{1 + (a_{nr}/a)^2}},
\end{equation}
Eq.(\ref{eq:veldm}) describes the exact velocity evolution of a decoupled massive particle. The transition between relativistic to non-relativistic is smooth and continuous, see \cite{Mastache:2019bxu} for a generalize transition. This evolution is general and valid for any massive decoupled particles (WDM, CDM or massive  neutrinos). If $a\gg a_{nr}$ it is clear that Eq.\eqref{eq:veldm} reduce to the non-relativistic limit where $v_{nr}(a) \sim a_{nr}/a$.

\begin{figure}[t]
\centering 
    \includegraphics[width=0.75\textwidth]{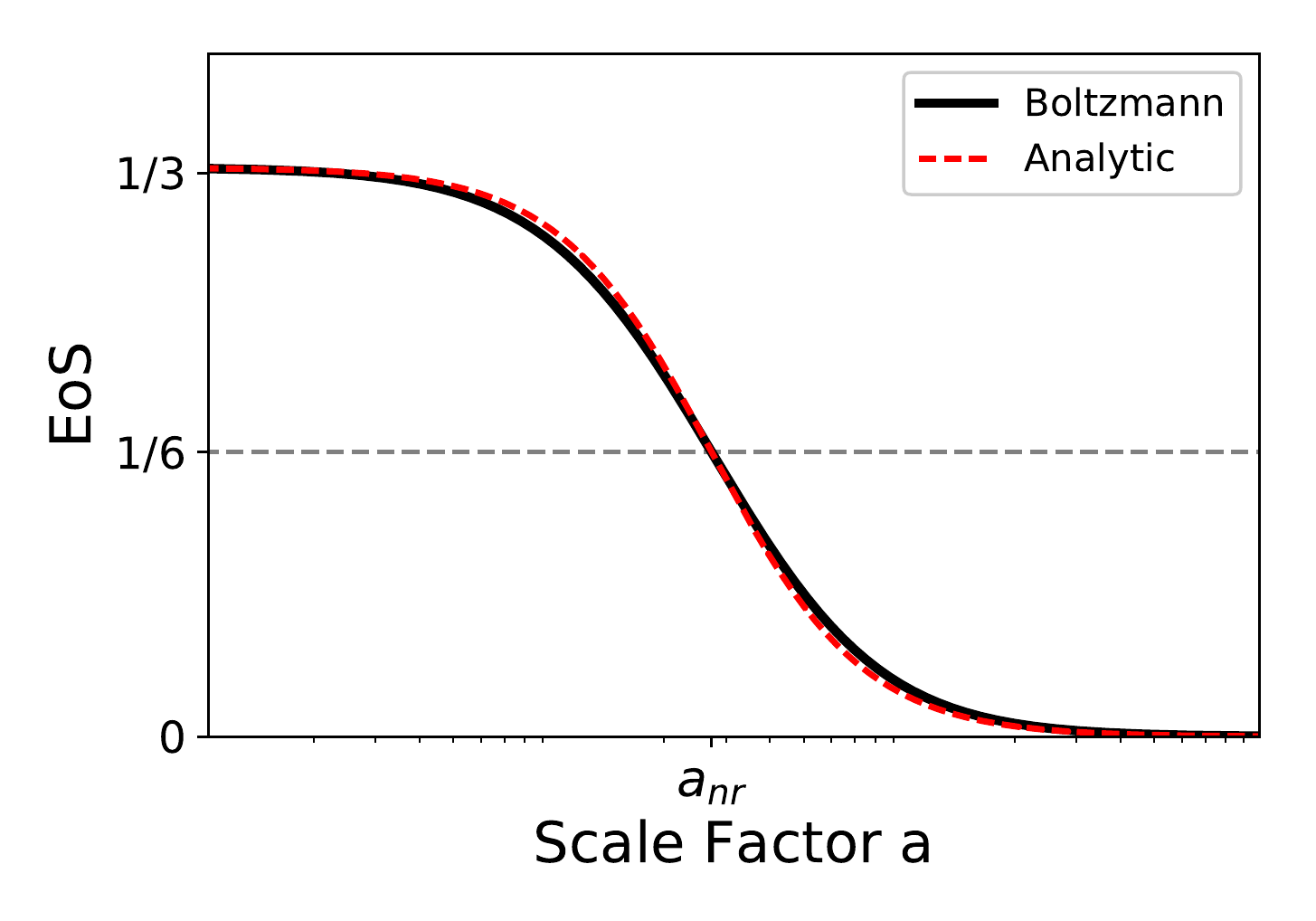}
    \caption{\footnotesize{ Plot of the equation of state for a non-cold dark matter. The continuous black line is obtained from solving Boltzmann equations using CLASS, the red dashed line is the analytic expression for the EoS of WDM, Eq.\eqref{eq:veldm} along with Eq.\eqref{eq:eos}.  }}
    \label{fig:eos} 
\end{figure}

The pressure of any generic particle is given by $P=\langle |\bar p|^2 \rangle n / 3\langle E \rangle$ and the energy density is given by $\rho= \langle E\rangle\, n$, with $n$ being particle number density, $\langle|\bar p|^2\rangle$ is the average quadratic momentum and $\langle E\rangle$ the average energy of the particles. Therefore the equation of state (EoS), $\omega = P/\rho$, is given by
\begin{equation}\label{eq:eos}
    \omega = \frac{ \langle |\bar p|^2\rangle}{3\langle E\rangle^2}= \frac{v(a)^2}{3} \, .
\end{equation}
We plot $\omega_{\rm bdm}$ in Fig.\ref{fig:eos}.
The EoS of DM have been investigates using the CMB and large scale structure (LSS) \cite{Muller:2004yb}, and gravitational lensing data. \cite{Faber:2005xc, Serra:2011jh} confirming that DM should be cold when strucuture began to cluster. We integrate the continuity equation, $\dot\rho=-3H(\rho+P)$, using Eq.(\ref{eq:eos}) to obtain the analytic evolution of  the background  $\rho_{\rm bdm}(a)$. For all $a$ we have,

\begin{equation}
    \rho_{\rm dm}(a) = \rho_{\rm dmo} \left( \frac{a}{a_{o}} \right)^{-4}\left( \frac{v_{o}}{v(a)} \right)  \label{eq:rho_bdm}  
\end{equation}  
with
\begin{equation}
    \frac{v_{o}}{v(a)} =  \frac{a}{a_o}\left(  \frac{1 + \left(\frac{a_{nr}}{a}\right)^2}{1 + \left(\frac{a_{nr}}{a_{o}}\right)^2}  \right)^{1/2}     \; .
\end{equation}  
When $a_{nr} \ll a_o$ and $a_{nr} \ll a$ we have
\be\la{f}
  \frac{v_o}{ v(a)} \sim \frac{a}{a_o} ,
\ee
then, the fluid behave as matter. Moreover, when $a_{nr} \ll a_o$ and $a \ll a_{nr}$ we have 
\be\la{f}
  \frac{v_o}{ v(a)} \sim \frac{a_{nr}}{a_o},
\ee
since the last quantity is a constant the fluid behaves as radiation. As seen in Eq.\eqref{eq:veldm}) a massive particle (WDM or CDM) becomes non-relativistic at $a_{nr}$ with $v(a_{nr})=1/\sqrt{2}$ and has only one free parameter, the scale factor $a_{nr}$. The density evaluated at $a_{nr}$ is given by $\rho_{\rm dm}(a_{nr}) \simeq \sqrt{2} \rho_{\rm dmo}(a_{nr}/a_o)^{-3} $. Therefore the physics of any massive particle with smooth continuous velocity transition can be described by $a_{nr}$ or the mass of the particle, i.e. $m_{\rm wdm}$. 

More complex approaches have been studied \cite{Hu:1998kj, Kopp:2016mhm, Thomas:2016iav, Kopp:2018zxp}
where they toke generalized properties of DM such as the sound speed and viscosity and put constrains with observational datasets. 
 
%%%%%%%%% FIG :: EOS %%%%%%  

% que corresponde a tantas partículas fermiónicas,  que puede elacionarse a 6 partículas fermionicas.

%%%%%%%%%%%%%%%%%%%%%%%%
%%%  SUB-SECTION  FREE-STREAMING %%%
\subsection{$m_{\rm wdm} - a_{nr} - T_{\rm wdm} $ relation}\label{ssec:manr_relation}
%%%%%%%%%%%%%%%%%%%%%%%%
Several constraints has been placed around the mass of the WDM based on different methods then it would be useful to have a relation between the mass ($m_{\rm wdm}$) and the time when DM become non relativistic ($a_{nr}$). Among current constrains on the mass of WDM are the ones based on the abundance of redshift $z = 6$ galaxies in the Hubble Frontier Fields, $m_{\rm wdm} > 2.4$ keV \cite{Menci:2016eui}. Based on galaxy luminosity function at $z \sim 6 - 8$, $m_{\rm wdm} > 1.5$ keV  \cite{Corasaniti:2016epp}. Using lensing surveys such as CLASH, $m_{\rm wdm} > 0.9$ keV \cite{Pacucci:2013jfa}. The upper limit of the mass is given by the high redshift Ly-$\alpha$ forest data which put lower bounds of $m_{\rm wdm} > 3.3 keV$  \cite{Viel:2013apy}, however hydrogen gas may not accurately trace the distribution of dark matter and the lower limit for the mass is still unsettled (see more details in \cite{Hui:2016ltb}).

We can then write the energy density as
\begin{eqnarray}
   \rho(T) &\propto& \frac{\pi^2 g_{\rm wdm}}{30}  T^4   \\   
   \rho(T_{nr}) &=& \frac{\pi^2 g_{\rm wdm}\beta^4}{30}  m_{\rm wdm}^4
\end{eqnarray}\label{eq:rho_simply}
where the constant $\beta$ take into account the relation between the mass $m_{wdm}$ and its temperature $T$, possible high order terms for the energy density as function of the temperature, as well as the statistical nature of the DM particle. For ultra-relativistic fermion particles the constant $\beta$ is given by (see appendix \ref{app:rho})
\begin{equation}
    \beta^4 =  \frac{2\sqrt{2}}{63\pi^4}  \left( \frac{7\pi^4}{9} - 60\sqrt{2}\xi(3) + 5 \pi^2 \right) \; ,
\end{equation} \label{eq:beta}
this is, $\beta \simeq 0.32$.

\begin{figure}[t]
\centering 
    \includegraphics[width=0.75\textwidth]{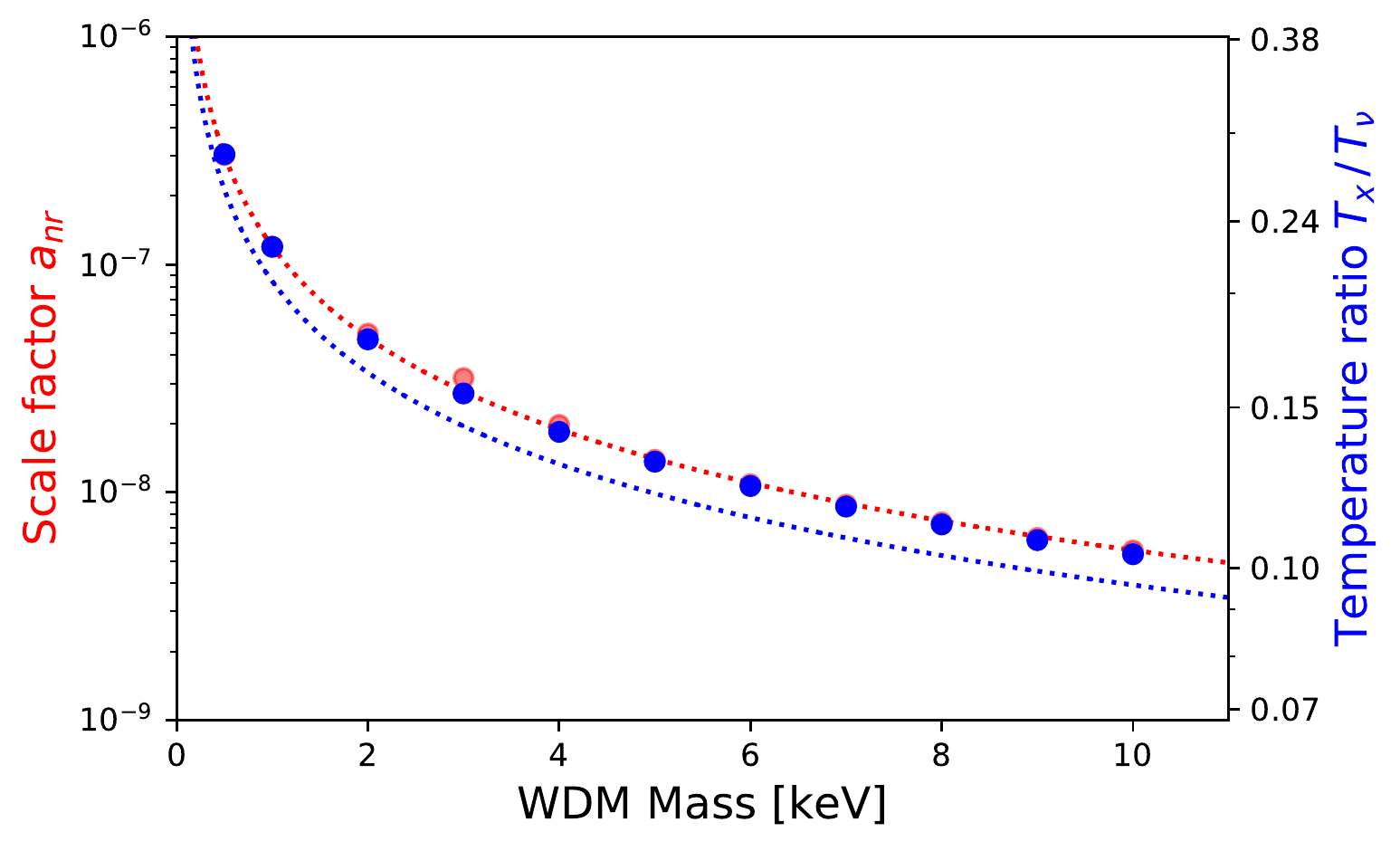}
    \caption{\footnotesize{ Plot of the WDM non-relativistic transition, $a_{nr}$ and temperature ratio $T_x/T_\nu$ as a function of the WDM mass. Red dots are the value obtained from the CLASS code, red dotted line is the theoretical prediction given by Eq.\eqref{eq:anrao}. Blue dots are the value of the temperature ratio predicted by Eq.\eqref{eq:ratio_T_viel} in  \cite{Viel:2005qj, Bode:2000gq}. Blue dotted line is the theoretical prediction by Eq.\eqref{eq:ratio_T}. }}
    \label{fig:anr_TxTv} 
\end{figure}

The WDM particle when $a \approx a_{nr}$ can be computed as a function of the temperature given by Eq.\eqref{eq:rho_simply} $\rho(T_{nr}) =(\pi^2 g_{\rm wdm}\beta/30)  m_{\rm wdm}^4$, this must me the same to the evolution of the energy density as function of the scale factor, $\rho(a_{nr}) \simeq \rho(T_{nr})$, this is,  $\rho_{dmo}(a_o/a_{nr})^3 \sqrt{2}=  \beta (\pi^2 g_{\rm wdm}/30)  m_{\rm wdm}^4$. Assuming a value for $g_{\rm wdm}=7/4$ we solve for $a_{nr}$
% Sacar la equacion de la beta
\be
\fr{a_{nr}}{a_o} =  2.77  \times 10^{-8}   \le(\fr{\omega_{dmo}}{0.120}\ri)^{1/3} \le(\fr{3\,keV}{m_{\rm wdm}}\ri)^{4/3}  \, ,
\la{eq:anrao}\ee
where $\omega_{dmo}$ is the amount of DM today. If we fit $\beta$ to the value of $a_{nr}$ obtained from CLASS we found that the best fit is $\beta = 0.34$ which is only 6\% from the theoretical value Eq.\eqref{eq:beta}, which reflect that $T = 3 m$ at $a = a_{nr}$. With the numerical code CLASS we obtained the EoS for a 3 keV WDM particle and look for the time where it becomes non-relativistic, to find that $a_{nr}^{\rm class}  = 3.18 \times 10^{-8}$,  just a $14.8\%$ different with the expected value, which is the largest difference we obtained in the mass range of 0.5 -10 keV, the average difference in the same mass range is 2.6\% as can bee seen in Fig.\ref{fig:anr_TxTv}.

%%%%%%%%% FIG :: EOS %%%%%%  

We can relate the time when two  different WDM become non-relativistic from Eq.(\ref{eq:anrao}) and  we find
\begin{eqnarray}
a_{nr} &=& a_{nr}^\prime \le( \fr{m_{\rm wdm}^\prime}{m_{\rm wdm}}\ri)^{4/3}. \la{mm}  \\
a_{nr} &=&  1.20\times10^{-7} \le( \fr{1 \; {\rm keV}}{m_{\rm wdm}}\ri)^{4/3}.
\end{eqnarray}

In Table \ref{tab:table1} we show so me WDM cases ($m_{\rm wdm} = {1, 3, 10}$ keV) in which we compare the numerical results for $a_{nr}$ obtained from this analytical calculations and the ones obtained from solving the Boltzmann equations using the numerical code CLASS for a non-cold DM.

When computing the Boltzmann equation the temperature relative to neutrinos or photons is also important. We can compute the WDM temperature, $T_x$, in the same fashion we compute Eq.\eqref{eq:anrao}. After we have  $\rho(a_{nr}) = \rho(T_{nr})$ we take the ratio with respect the neutrino temperature, $T_\nu$, at $a=a_{nr}$ obtaining 
\begin{eqnarray}
   \left( \frac{T_{x}}{T_\nu} \right)^4 &\simeq& \sqrt{2} \frac{g_\nu}{g_{\rm wdm}} \frac{\omega_{dmo}}{\omega_{\nu o}} \frac{a_{nr}}{a_o}   \\
    \frac{T_{x}}{T_\nu} &=& 0.1480 \le(\fr{\omega_{dmo}}{0.120}\ri)^{1/3} \le(\fr{3\,keV}{m_{\rm wdm}}\ri)^{1/3}   \label{eq:ratio_T}
\end{eqnarray}
where we use Eq.\eqref{eq:anrao}, assume massless neutrinos, and we take $\sigma/\beta \simeq 1$. The difference between this equation and previous results, \cite{Viel:2005qj, Bode:2000gq}, is around 8\% for all cases, therefore the difference can be parametrized. Using the $m_{\rm wdm} = 1$ keV as a pivot case, from Eq.\eqref{eq:ratio_T} we find that
\begin{equation}
     \frac{T_{x}}{T_\nu} = 0.2062 \left(  \frac{\rm 1 \; keV}{m_{\rm wdm}} \right)^{1/3}
\end{equation}

In \cite{Viel:2005qj, Bode:2000gq} previously found the temperature ratio by fitting several numerical simulations and is given by
\begin{equation}
     \frac{T_x}{T_\nu} = 5.5813 \left( \frac{\alpha}{Mpc} \right)^{1.205} \left( \frac{m_{\rm wdm}}{keV} \right) \left( \frac{\omega_{\rm wdm}}{0.122}  \right)^{0.193} \label{eq:ratio_T_viel}
\end{equation}
where the $\alpha$ parameter is
\be \label{eq:alpha}
\alpha= 0.049 \le( \fr{m_{\rm wdm}}{1 keV} \ri)^{1.11}   \le( \fr{\Omega_{\rm wdm}}{0.25} \ri)^{0.11}  \le( \fr{h}{0.7} \ri)^{1.22}  h^{-1} Mpc \, .
\ee
Perhaps is not a clear relation between Eq.\eqref{eq:ratio_T} and \eqref{eq:ratio_T_viel} however their evaluations proves they give the similar numerical results and even more, Eq.\eqref{eq:ratio_T} gives a more clear way to understand the physics of the WDM.

%%%%%%%% FIG :: MPS BDM %%%%%%
%%%%%%%% FIG :: MPS BDM %%%%%%
\begin{figure}[t]
\centering 
     \includegraphics[width=0.75\textwidth]{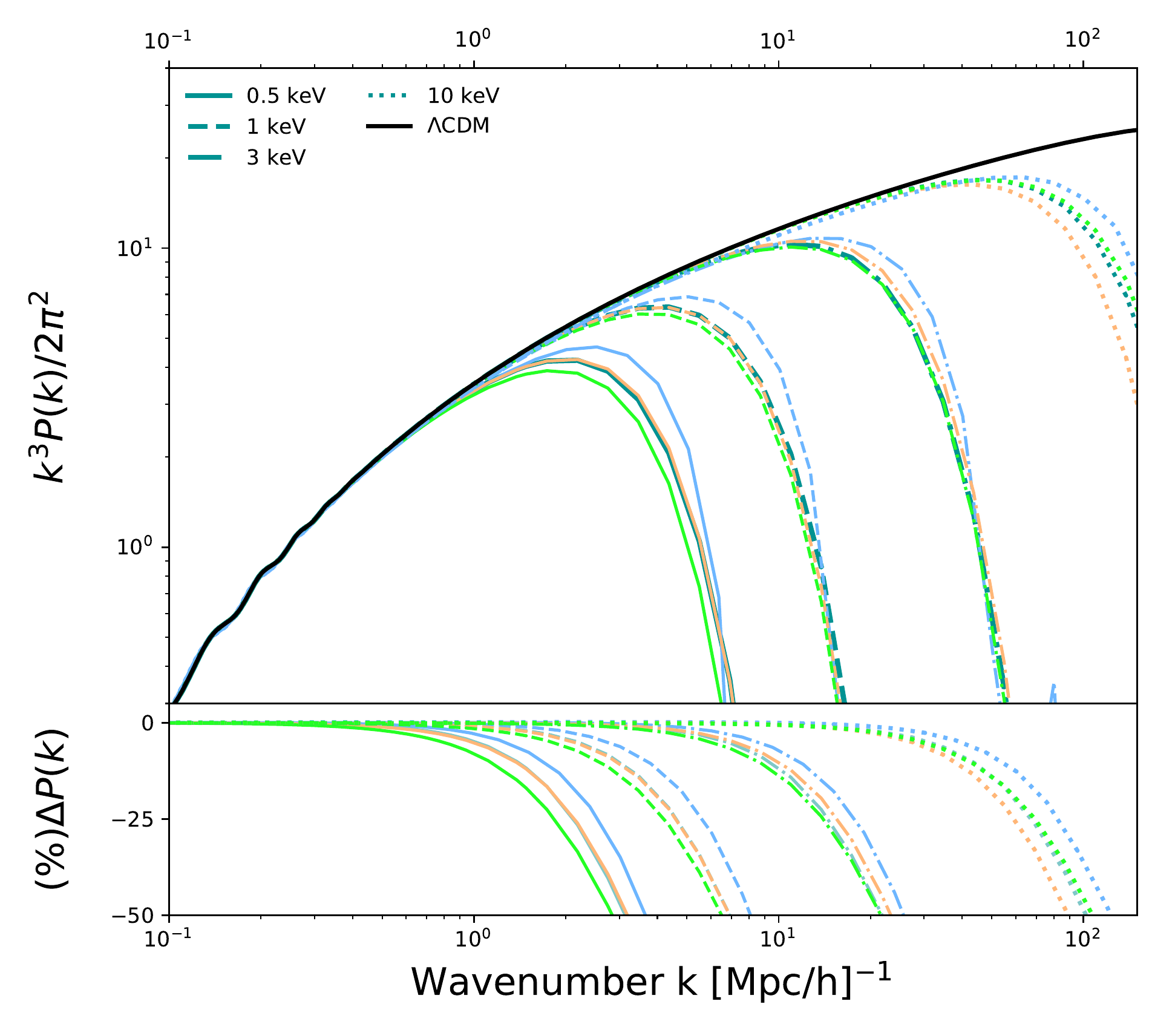} 
      \caption{\footnotesize{ {\bf Top panel}. Plots of linear dimensionless matter power spectra for the CDM (black solid line) and WDM for different mass values. 
      Line styles represent different mass, 0.5,1, 3, and 10 keV for straight, dashed, dot-dashed, and dotted lines, respectively. 
      Different color represent different approaches, dark green lines are those obtained from CLASS,  blue lines is the fluid approximation  (Sec.\ref{ssec:pert}), 
      orange and green lines is the matter power spectrum obtained from the standard and virial transfer function (Eqs. \eqref{eq:transf_virial} and \eqref{eq:transfer_viel}). 
      {\bf Bottom panel}. We show the percentage difference, $\Delta P(k)$, between CDM and the one obtained with WDM. Notice that $k_{1/2}$ is defined when the difference 
      between different matter power spectrum reaches a 50\% difference.  }}  \label{fig:mps}
\end{figure}

%%%%%%%%%%%%%%%%%%%%%%%%
%%%              SECTION      LSS               %%%
\section{Large Scale Structure in BDM scenario}\label{sec:lls}
%%%%%%%%%%%%%%%%%%%%%%%%
In order to compute the cut-off scale, we compute the Boltzmann equations for WDM, using the fluid approximation in CLASS. But first, we show the equation where we can compute the matter power spectrum and WDM temperature ratio with some fitting formulas. 

\subsection{Perturbations}\label{ssec:pert}

\begin{table}[t]
	\centering     
	\begin{tabular}{l|ccc}
		\hline
	 	& \multicolumn{3}{c}{TT,TE,EE+lowE+WiggleZ}  \\
		Parameter & Best fit & 68\% limits & 95\% limit    \\
		\hline
		$\Omega_{b} h^2$  & 0.02223 & $0.02233\pm 0.00015$ & $0.02233\pm 0.00030$  \\
		$\Omega_{cdm}h^2$  & 0.1199 & $0.1187\pm 0.0013$ & $0.1187\pm 0.0026$  \\
		$100\theta_{s}$  & 1.04190 & $1.04180\pm 0.00031$ & $1.04180\pm 0.00062$  \\
		$\tau_{\rm reio}$  & 0.0487 & $0.0681_{-0.0176}^{+0.0133}$ & $0.0681_{-0.0281}^{+0.0268}$  \\
		$\ln(10^{10}A_{s})$  & 3.032 & $3.068_{-0.034}^{+0.026}$ & $3.068_{-0.055}^{+0.057}$  \\
		$n_{s}$  & 0.9588 & $0.9659\pm 0.0045$ & $0.9659\pm 0.0093$  \\
		\hline
		$a_{nr}$ & $3.1158\times10^{-6}$ & $<4.3205\times10^{-6}$ & $<8.9057\times10^{-6} $
	\end{tabular}
	\caption{In this table we show the time when a WDM stop being relativistic, $a_{nr}$, the free streaming scale, $\lambda_{fs}$ [Mpc/h] from Eq.\eqref{eq:lambda_fs}, 
	the correspondent mode $k_{fs}$ [h/Mpc] and Jeans mass, $M_{fs}$ $[M_\odot / h^3]$ for different masses $m_{\rm wdm} = {1,3,10}$ keV. 
	For each case we show the $a_{nr}$ obtained from solving the Boltzmann equations using CLASS, and the one obtained from the analytic expression Eq.\eqref{eq:anrao}.  }
	\label{tab:mcmc_results}
\end{table}

We follow \cite{Ma:1995ey} to compute the fluid limit to the perturbed equations in k-space in the synchronous gauge for the WDM are

\begin{align}
    \dot{\delta_c} &= -\left(1+\omega\right)\left(\theta + \frac{\dot{h}}{2} \right) -  2\omega H \delta_c \frac{1-3\omega}{1+\omega}   \label{delta_mat} \\[1ex]
    \dot{\theta} &= -H\theta \frac{ (1 - \omega)(1-3\omega) }{ 1+\omega } + k^2\delta \, \frac{\omega(5-3\omega)  }{ 3(1+\omega)^2} - k^2\sigma  \\[1ex]
    \dot { \sigma } &=
           - 3  \frac { \sigma } { \tau } -  2 \sigma H  \left[ \frac{1-3\omega}{1+\omega}  \right]  + 
           \frac { 8 } { 3 } \left( \theta + \frac{\dot { h }}{2} \right) \frac{\omega^2 (5 - 3\omega)}{  \left( 1+\omega \right)^2} 	  \label{eq:sigma_matter}
\end{align}

where $\delta$ is the contrast, $\theta$ is the divergence of the velocity in Fourier space and $\sigma$ is the anisotropic stress perturbations. The dot represents the derivative respect to the conformal time, $\tau \equiv \int d t/a(t)$, $H$ is the Hubble parameter.

In Eq.(\ref{eq:sigma_matter}) we have taken the anisotropic stress approximation for neutrinos \cite{Hu:1998kj, Lesgourgues:2011rh} and ignore the $\dot{\eta}$ term that slightly the computation of the matter power spectrum \cite{Lesgourgues:2011rh}. We have also used the relation ${\dot \omega} = -2 H \omega (1-3\omega)$.

For the numerical computations for the matter power spectrum we adopt Planck 2018 cosmological results \cite{Aghanim:2018eyx}. We adopt a flat Universe with $\omega_{dmo} = 0.12$, and $\omega_{bo} = 0.02237$ as the CDM matter and baryonic omega parameter. $h = 0.6736$ is the Hubble constant in units of 100 km/s/Mpc, $n_s = 0.965$ is the tilt of the primordial power spectrum. $z_{\rm reio} = 7.67$ is the redshift of reonization and $\ln(10^{10} A_s) = 3.044$, where $A_s$ is the amplitud of primordial fluctuations.

In Fig.\ref{fig:mps} we show the dimensionless matter power spectrum obtained with CLASS code \cite{Blas:2011rf} taking into account WDM fluid approximation, Eqs.(\ref{delta_mat})-(\ref{eq:sigma_matter}). We show the matter power spectrum for different values of $m_{\rm wdm}  = \{0.5, 1, 3, 10 \}$ keV. The bigger the mass the colder the DM is, therefore for bigger masses the difference with CDM decreases. In Fig.\ref{fig:mps} blue lines are the matter power spectrum obtained for different WDM masses using the fluid approximation equations. The percentage difference between $\Lambda$CDM different approaches is shown in the bottom panel of Fig.\ref{fig:mps}. 

The effect of the free-streaming is to suppress structure formation below a threshold scale, therefore the matter power spectrum shows a cut-off at small scales depending on the value of $m_{\rm wdm}$, equivalently $a_{nr}$. The smaller the scale of the transition $a_{nr}$, the bigger the mass for the WDM particle is, and the power is damped at bigger scales. Transitions of the order to $a_{nr} \lesssim 10^{-8}$ WDM is indistinguishable from CDM at observable scales, $k \sim \, \mathcal{O} (10) \; {\rm h \; Mpc^{-1}}$, this corresponds to $m_{\rm wdm} \gtrsim 3$ keV. 

The CMB power spectrum can also be computed, but the difference with respect to the fiducial $\Lambda$CDM model is barely perceptible one can notice an increased the height of the acoustic peaks of less than 1\% because difference respect CDM, this increment is the free-streaming also increase the acoustic oscillations.

Our approach given a deeply insight on the nature of WDM, even more, another feature of using the fluid approximation is the CPU time when solving Eqs.(\ref{delta_mat})-(\ref{eq:sigma_matter}). Each full solution to the fluid approach equations can be, at least 8x faster than the solution for the full Boltzmann equations, the last approach can even be slower depending the computational parameters specified in the code, see \cite{Lesgourgues:2011rh} for full details. As a direct consequence of using a faster computation using the fluid approximation we can compute marginalized values and confidence regions for cosmological parameters for the $\Lambda$WDM model. In Table \ref{tab:mcmc_results} we show the results obtained combining Planck TT,TE,EE + lowE \cite{Aghanim:2018eyx} and WiggleZ \cite{Parkinson:2012vd} data sets. We notice that only Planck data is not able to constrain $a_{nr}$. 

%%%%%%%%    FIG:MCMC  %%%%%%%%%
\begin{figure}[t]
  \centering
    \includegraphics[width=0.85\textwidth]{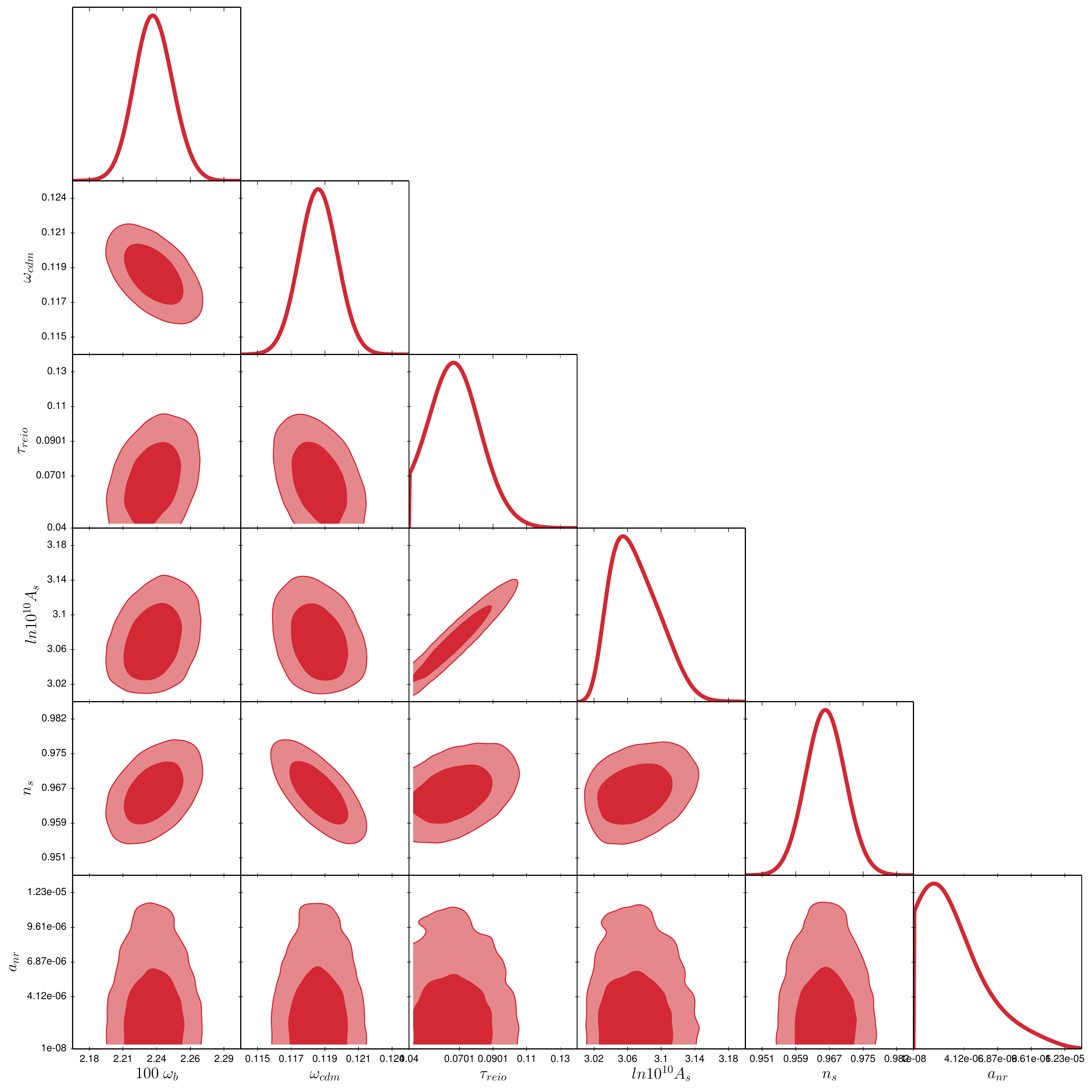}
    \caption{\footnotesize{
        Constraints on parameters of the $\Lambda$WDM using the fluid approximation model using Planck EE, TE,TT+lowE combined with WiggleZ data set. Parameters on the bottom axis are our sampled MCMC parameters with flat priors. Contours contain 68\% and 95\% of the probability.  }}
    \label{fig:mcmc_results}
\end{figure}

In Fig. \ref{fig:mcmc_results} and Table \ref{tab:mcmc_results} we show the 86\% and 95\% confidence level for the Vanilla parameters and the non-relativistic transition scale $a_{nr}$. First notice that all parameters are consistent at 85\% confidence level with latest Planck results using TT,TE,EE+lowE data sets. The key parameter to our approach is $a_{nr}$, from Montearlo-Markov chains (Fig. \ref{fig:mcmc_results}) we can conclude that the parameters is not degenerated with other parameter, and only data for the matter power spectrum can constrain its value. We are only able to put lower constrains to $a_{nr}$, equivalently to the lower value of $m_{\rm wdm}$, which makes sense, because the smaller the value for $a_{nr}$ is the bigger is the mass for the DM particle, this is, a physics closer to CDM.

The lower bounds to $a_{nr}$ are:
\begin{eqnarray*}
     a_{nr} < 4.3205\times10^{-6}; \;\;\; m_{\rm wdm} > 70.3 \; {\rm eV}    \;\;\;  {\rm for \;} 86\% \; {\rm c.l.}  &&\\
     a_{nr} <  8.9057\times10^{-6}; \;\;\; m_{\rm wdm} > 40.9 \; {\rm eV}    \;\;\;  {\rm for \;} 95\% \; {\rm c.l.} &&\; ,
\end{eqnarray*}
where c.l. stand for confidence level. The $k_{1/2}$ characterize the scale of the cut-off in the matter power spectrum, it is defined as the point where the ratio of the power spectrum $P^{\rm wdm}/P^{\rm cdm} = 0.5$. The $86\%$ confidence constrain the WDM mass to $m_{\rm wdm}^{\rm 1\sigma} = 70.3$ keV, this WDM mass has a cut-off in the matter power spectrum at a wavenumber scale of $k_{\rm 1/2} = 0.28 \; {\rm h \; Mpc^{-1}}$, which is close to the highest observation value given by WiggleZ, $k_{\rm max} \approx 0.4  \; {\rm h \; Mpc^{-1}}$. Moreover, $m_{\rm wdm}^{\rm 1\sigma}$ gives a free-streaming mass $M_{fs} = 1.61 \times 10^{13} \; {\rm M_\odot/h^3}$, a very high value which is not consistent with current large scale structure observation, however, no data set to constrain the free-streaming mass was included in our MCMC analysis. Notice that a $m_{\rm wdm} \approx 7$ keV would have $M_{fs} \simeq 10^6  \; {\rm M_\odot/h^3}$ which would be consistent with the lowest mass of Milky Way satellite galaxies, but this heavier WDM particle have a cut-off in the matter power spectrum at a wavenumber $k_{\rm 1/2} = 69.1 \; {\rm h \; Mpc^{-1}}$, a very low scales, out of reach of current matter power spectrum observations. 

The constraints values are lower than constrains based on the abundance of redshift $z = 6$ galaxies in the Hubble Frontier Fields, whose constrains is $m_{\rm wdm} > 2.4$ keV \cite{Menci:2016eui}. Galaxy luminosity studies at $z \sim 6 - 8$ put constrains on $m_{\rm wdm} > 1.5$ keV  \cite{Corasaniti:2016epp}. Lensing CLASH survey provide $m_{\rm wdm} > 0.9$ keV lower bounds \cite{Pacucci:2013jfa}. High redshift Ly-$\alpha$ forest data put lower bounds of $m_{\rm wdm} > 3.3 keV$ \cite{Viel:2013apy}. Therefore, we conclude that, despite our MCMC analysis is consistent with current matter power spectrum observations from WiggleZ, more observations at small scales is needed to constrain the value of $a_{nr}$, however, at smallest scales means bigger wavenumber values, where linear perturbation is no longer valid and a non-linear study, out of the scope of this paper, needs to be done.

\subsection{Transfer Functions} \label{ssec:transfer_func}
 The matter power spectrum can also be obtained from a parametrized transfer function defined as $T^2 = P^{\rm wdm}(k)/P^{\rm cdm}(k)$. It has been proposed \cite{MacorraVirial:2019gui} a virial approach to the transfer function, $T$, as a function of the virial mode $k_v$  given by
\begin{equation}  \label{eq:transf_virial}
  T(k) =   \left(1 + \left(\frac{k}{k_{v}} \right)^{2\beta}\right)^{-9/\beta}
\end{equation}
with $\beta = 1.02$. All the physical properties of WDM are imprinted in the virial mode $k_v$ which is directly  related to the  free-streaming scale as $k_v = 2k_{fs}$,  where the free streaming scale is defined as $k_{fs}=2\pi/\lm_{fs}$ where  $\lambda_{fs} = \int dt v(a)/a(t)$, using Eq.\eqref{eq:veldm} it has been proved  \cite{MacorraVirial:2019gui} that 

%%% Ejemplificar con valor de lambda_fs y k_fs de la Eq.(15)
\begin{table}[t]
	\centering     
	\begin{tabular}{l|cccc}
		Model & $a_{nr}$ & $\lambda_{fs}$ & $k_{fs}$ & $M_{fs}$ \\
		\hline
		1 keV - Boltzmann &   $1.20\times 10^{-7}$        & 0.398 & 15.79   & $1.10\times 10^{9}$ \\
		1 keV - analytic &        $1.20\times 10^{-7}$        & 0.399 & 15.74   & $1.10\times 10^{9}$ \\
		3 keV - Boltzman &      $3.18\times 10^{-8}$       & 0.120 & 52.48   & $2.99\times 10^{7}$ \\
		3 keV - analytic &        $2.77\times 10^{-8}$        & 0.106 & 59.39   & $2.06\times 10^{7}$ \\
		10 keV - Boltzmann & $5.56\times 10^{-9}$        & 0.024 & 259.7 &  $2.47\times 10^{5}$ \\
		10 keV - analytic &      $5.57\times 10^{-9}$        & 0.024 & 259.3 &  $2.48\times 10^{5}$ \\
	\end{tabular}
	\caption{In this table we show the time when a WDM stop being relativistic, $a_{nr}$, the free streaming scale, $\lambda_{fs}$ [Mpc/h] from Eq.\eqref{eq:lambda_fs}, 
	the correspondent mode $k_{fs}$ [h/Mpc] and Jeans mass, $M_{fs}$ $[M_\odot / h^3]$ for different masses $m_{\rm wdm} = {1,3,10}$ keV. 
	For each case we show the $a_{nr}$ obtained from solving the Boltzmann equations using CLASS, and the one obtained from the analytic expression Eq.\eqref{eq:anrao}.  }
	\label{tab:table1}
\end{table}

\begin{equation}\label{eq:lambda_fs}
   \lambda_{fs} = 0.011 \left( 8.82 + \left( \frac{1+3411}{1+z_{eq}} \right) \left( \frac{2.77\times10^{-8}}{a_{nr}} \;. \right) \right) \frac{\rm Mpc}{\rm h}  
\end{equation}

The standard parametrization is given by \cite{ Viel:2005qj, Bode:2000gq, Colin:2000dn}
\begin{equation}\label{eq:transfer_viel}
  T  ( k )  = \left[ 1 + ( \alpha k ) ^ { \beta} \right] ^ { \gamma }, 
 \end{equation}
 where the parameter $\alpha$ is defined in Eq.\eqref{eq:alpha} and $\beta = 2\nu$, $\gamma = -5/\nu$ with $\nu = 1.12$. We compare the matter power spectrum obtained from the virial and standard transfer function, the one obtained from the full Boltzmann solution using CLASS and the fluid approximation in Fig.\ref{fig:mps}.

%%%%%%%%%%%%%%%%%%%%%%%%%%
%%% SUB-SECTION PRESS-SCHECHTER %%%
\subsection{Halo Mass Function}\label{ssec:preschechter}
%%%%%%%%%%%%%%%%%%%%%%%%%%
The change in the matter power spectrum is known to strongly affect large scale structure, we compute the the abundance of structure using the PressSchechter approach \cite{Press:1973iz}. With the linear matter power spectrum (see Sec.\ref{sec:lls}) as an input we  compute the halo mass function as
\begin{equation}
  \frac{dn}{d\log M} = M \frac{dm}{dM} = \frac{1}{2} \frac{\overline{\rho}}{M} \mathcal{F}(\nu) \frac{d \log \sigma^2}{ d \log M}
\end{equation}
where $n$ is the number density of haloes, $M$ the halo mass and the the peak-height of perturbations is given by
\begin{eqnarray}
   \nu = \frac{\delta_c(z) }{\sigma(M)}, 
\end{eqnarray}
where  $\delta_c = 1.686$ is the overdensity required for spherical collapse model in a $\Lambda$CDM cosmology. The average density is $\overline{\rho} = \Omega_m \rho_c$, where $\rho_c$ is the critical density of the Universe. Here $\Omega_m = \Omega_{c} + \Omega_b $. The variance of the linear density field  on mass-scale, $\sigma^2(M)$, can be computed from the following integrals
\begin{equation} \label{eq:sigma_8}
  	\sigma^2(M) = \int_0^\infty dk \frac{k^2 P_{\rm lin}(k)}{2\pi^2}| W(kR) |^2.
\end{equation}

%%%%%%%%    FIG:PRESS-SCHECHTER  %%%%%%%%%
\begin{figure}[t]
	  \centering
	  \includegraphics[width=0.75\textwidth]{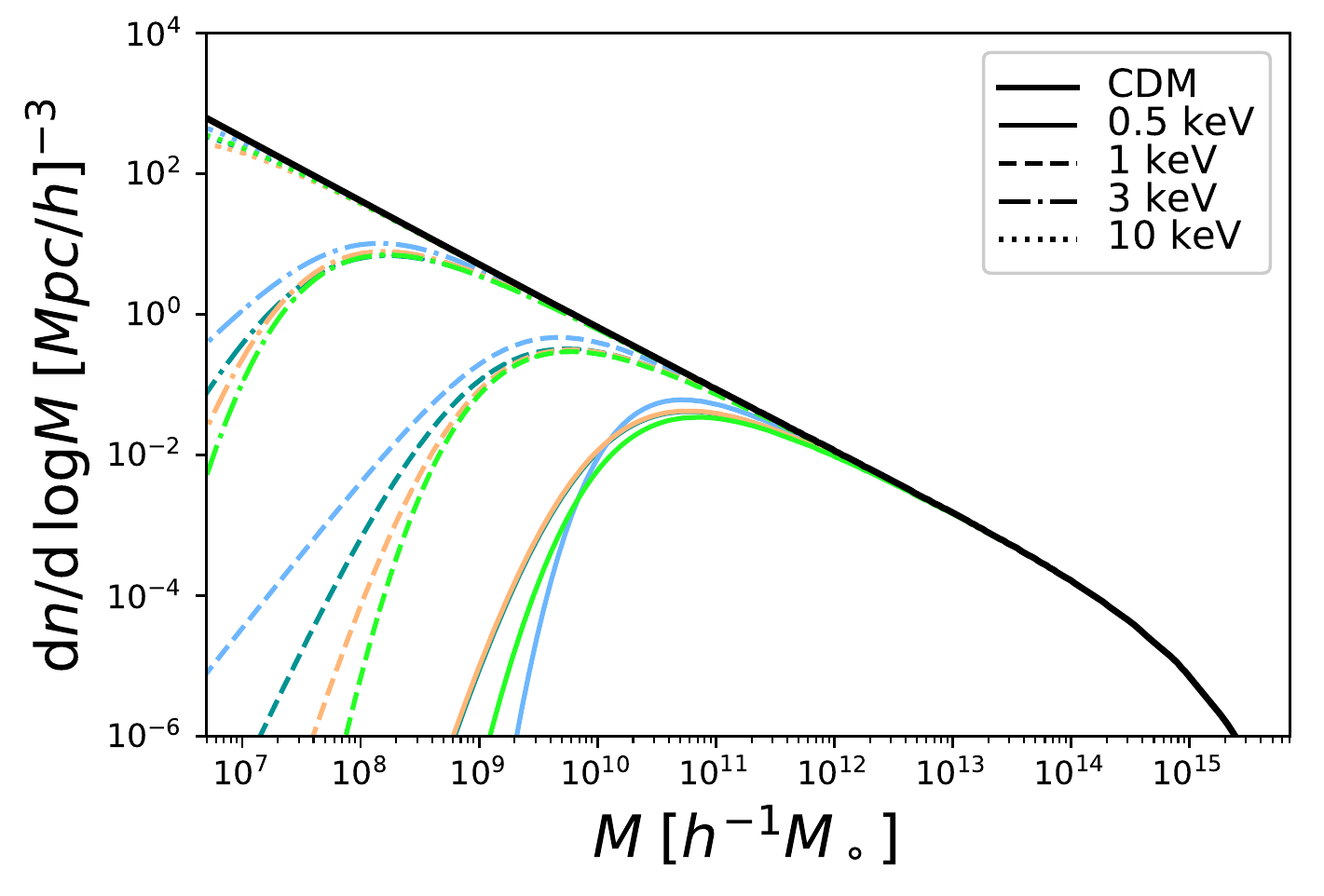}
	  \caption{\footnotesize{ Halo mass function as a function of halo mass computed using different  approaches for the matter power spectrum. 
	  The black solid line is the CDM model; Colors with line styles straight, dashed, dot-dashed, and dotted lines represent 0.5, 1, 3, and 10 keV WDM masses, respectively. 
	  Color represent, dark green is the CLASS solution, blue is the fluid approximation, orange is the standard transfer function, and light green the virial transfer function.   }}
	  \label{fig:pressschechter}
\end{figure}

Here we will use the sharp-k window function $W(x) = \Theta( 1 - kR)$, with $\Theta$ being a Heaviside step function, and $R = (3cM/4\pi \overline{\rho})^{1/3}$, where the value of $c = 2.5$ is proved to be best for cases similar as the WDM \cite{Benson:2012su}. Finally for the  first crossing distribution $\mathcal{F}(\nu)$ we adopt \cite{Bond:1990iw}, that has the form
\begin{equation}
   \mathcal{F}(\nu) = A\left( 1 + \frac{1}{\nu^{\prime p}} \right)\sqrt{ \frac{\nu^\prime}{2\pi} } e^{-\nu^\prime/2}
\end{equation}\label{eq:pressschechter}
with $\nu^{\prime} = 0.707\nu$, $p = 0.3$, and $A = 0.322$ is the normalization factor (ensuring that the integral $\int f(\nu)d\nu = 1$).  In Fig.\ref{fig:pressschechter} we show the halo mass function computed with the Press-Schechter approach and using different matter power spectrums, in particular the fluid approximation is plotted in blue.

%%%%%%%%%%%%%%%%%%%%%%%%
%%%        SECTION        RESULT            %%%
\section{Conclusion}\label{sec:conclusion}
%%%%%%%%%%%%%%%%%%%%%%%%
We have presented a generalization for the velocity dispersion of particles, Eq.\eqref{eq:veldm}, which is valid for any decoupled massive particle, generically known as non-cold DM, such as WDM and massive neutrinos. This velocity is a function of the scale factor and it also depends on the time the particle becomes non-relativistic, $a_{nr}$. The general framework for the velocity has some serious advantages that must be taken into account and can be enumerated: (1) Eq.\eqref{eq:veldm} embrace the concept that a particle stopped being relativistic when $p = m$, therefore it has been proven that one-to-one relation between $a_{nr}$ and the mass of the particle can be found, Eq.\eqref{eq:anrao}. (2) The velocity expression is simple and describes a smooth transition between relativistic to non-relativistic regimes fo the particle. (3) The continuity equation can be solved, Eq.\eqref{eq:rho_bdm}, and therefore is straight forward to compute the perturbation equations in the fluid approximation, which is terms of computational effort could save a significant amount of time. (4) A slight difference in the CMB power spectrum can be found, less than 1\%, but more importantly, the cut-off in the linear matter power spectrum (Fig.\ref{fig:mps}.) due to the free streaming, $\lambda_{fs}$, can be replicated it with great accuracy as well as in the halo mass function (Fig.\ref{fig:pressschechter}.), which is one of the most appealing features for WDM. (5) Using the velocity dispersion we can also compute the dark matter and neutrino temperature ratio, $T_x/T_v$ (Eq.\eqref{eq:ratio_T}), the free-streaming scale ($\lambda_{fs}$), and wavenumber ($k_{fs}$), which in turn, define the virial transfer function (\cite{MacorraVirial:2019gui}) that depends on the virial wavenumber as $k_v = 2 k_{fs}$, all these calculations only from the theoretical perspective which support a deeper understanding of the WDM physics and reproduce previous results with great accuracy.

We use the 3 keV WDM as a base example, with a solid theoretical analysis we compute that this particle becomes non-relativistic at $a_{nr} = 2.77 \times 10^{-8}$, 15\% above the value obtained from solving the Boltzmann equations encoded in CLASS. In general, the percentage difference between the numerical value computed from Boltzmann equations and the analytic one is on average 3\% in the mass range 1-10 keV.

We use the fluid approximation to compute 68\% and 95\% parameter contours from MCMC analysis from the CMB power spectrum and WiggleZ matter power spectrum, we put lower bounds to the mass $m_{\rm wdm} > 70.3$ eV within the 86\% c.l. We conclude that the analysis is consistent with WiggleZ matter power spectrum observations, however, more data or combination with other data sets that include large scale structure information at small scales need to be included to put stronger constrains to the $a_{nr}$ and therefore $m_{\rm wdm}$.

This framework, in which we have included the dispersion velocity of the WDM particle is a clear indication that we understand the background evolution and the fluid perturbation scheme of a WDM particle. This approach may be incorporated in a broad number of observational cosmological probes, Montecarlo analysis, theoretical analyzes, and N-body simulations, including forecasts for large scale structure measures, i.e. weak lensing  \cite{Markovic:2010te}, future galaxy clustering measures of the power spectrum \cite{vandenBosch:2003nk}. From future observation from large to small-scale clustering of dark and baryonic matter may be able to put more feasible constraints on $a_{nr}$ and therefore on the WDM mass, $m_{\rm wdm}$.

%%%%%%%%%%%%%%%%%%%%%%%%
%%%              SECTION      ACKNOWLEDGEMENTS               %%%
\section{ACKNOWLEDGEMENTS}\label{sec:acknowledgements}
%%%%%%%%%%%%%%%%%%%%%%%%
JM acknowledges Catedras-CONACYT financial support and MCTP/UNACH as the hosting institution of the Catedras program. AM acknowledge  support from PASPA-DGAPA, UNAM and Project IN103518 PAPIIT-UNAM. 

%\bibliographystyle{cas-model2-names}
\bibliographystyle{JHEP}
\bibliography{wdmfld}

\begin{thebibliography}{10}
\expandafter\ifx\csname url\endcsname\relax
  \def\url#1{\texttt{#1}}\fi
\expandafter\ifx\csname urlprefix\endcsname\relax\def\urlprefix{URL }\fi
\expandafter\ifx\csname href\endcsname\relax
  \def\href#1#2{#2} \def\path#1{#1}\fi

\bibitem{Aghanim:2018eyx}
N.~Aghanim, et~al., {Planck 2018 results. VI. Cosmological parameters}\href
  {http://arxiv.org/abs/1807.06209} {\path{arXiv:1807.06209}}.

\bibitem{Abbott:2016ktf}
T.~Abbott, et~al., {The Dark Energy Survey: more than dark energy ? an
  overview}, Mon. Not. Roy. Astron. Soc. 460~(2) (2016) 1270--1299.
\newblock \href {http://arxiv.org/abs/1601.00329} {\path{arXiv:1601.00329}},
  \href {http://dx.doi.org/10.1093/mnras/stw641}
  {\path{doi:10.1093/mnras/stw641}}.

\bibitem{Betoule:2014frx}
M.~Betoule, et~al., {Improved cosmological constraints from a joint analysis of
  the SDSS-II and SNLS supernova samples}, Astron. Astrophys. 568 (2014) A22.
\newblock \href {http://arxiv.org/abs/1401.4064} {\path{arXiv:1401.4064}},
  \href {http://dx.doi.org/10.1051/0004-6361/201423413}
  {\path{doi:10.1051/0004-6361/201423413}}.

\bibitem{Archidiacono:2013dua}
M.~Archidiacono, S.~Hannestad, {Updated constraints on non-standard neutrino
  interactions from Planck}, JCAP 1407 (2014) 046.
\newblock \href {http://arxiv.org/abs/1311.3873} {\path{arXiv:1311.3873}},
  \href {http://dx.doi.org/10.1088/1475-7516/2014/07/046}
  {\path{doi:10.1088/1475-7516/2014/07/046}}.

\bibitem{Kazantzidis:2003hb}
S.~Kazantzidis, L.~Mayer, C.~Mastropietro, J.~Diemand, J.~Stadel, B.~Moore,
  {Density profiles of cold dark matter substructure: Implications for the
  missing satellites problem}, Astrophys. J. 608 (2004) 663--3679.
\newblock \href {http://arxiv.org/abs/astro-ph/0312194}
  {\path{arXiv:astro-ph/0312194}}, \href {http://dx.doi.org/10.1086/420840}
  {\path{doi:10.1086/420840}}.

\bibitem{BoylanKolchin:2011de}
M.~Boylan-Kolchin, J.~S. Bullock, M.~Kaplinghat, {Too big to fail? The puzzling
  darkness of massive Milky Way subhaloes}, Mon. Not. Roy. Astron. Soc. 415
  (2011) L40.
\newblock \href {http://arxiv.org/abs/1103.0007} {\path{arXiv:1103.0007}},
  \href {http://dx.doi.org/10.1111/j.1745-3933.2011.01074.x}
  {\path{doi:10.1111/j.1745-3933.2011.01074.x}}.

\bibitem{Klypin:1999uc}
A.~A. Klypin, A.~V. Kravtsov, O.~Valenzuela, F.~Prada, {Where are the missing
  Galactic satellites?}, Astrophys. J. 522 (1999) 82--92.
\newblock \href {http://arxiv.org/abs/astro-ph/9901240}
  {\path{arXiv:astro-ph/9901240}}, \href {http://dx.doi.org/10.1086/307643}
  {\path{doi:10.1086/307643}}.

\bibitem{Garrison-Kimmel:2017zes}
S.~Garrison-Kimmel, et~al., {Not so lumpy after all: modelling the depletion of
  dark matter subhaloes by Milky Way-like galaxies}, Mon. Not. Roy. Astron.
  Soc. 471~(2) (2017) 1709--1727.
\newblock \href {http://arxiv.org/abs/1701.03792} {\path{arXiv:1701.03792}},
  \href {http://dx.doi.org/10.1093/mnras/stx1710}
  {\path{doi:10.1093/mnras/stx1710}}.

\bibitem{Sawala:2015cdf}
T.~Sawala, et~al., {The APOSTLE simulations: solutions to the Local Group's
  cosmic puzzles}, Mon. Not. Roy. Astron. Soc. 457~(2) (2016) 1931--1943.
\newblock \href {http://arxiv.org/abs/1511.01098} {\path{arXiv:1511.01098}},
  \href {http://dx.doi.org/10.1093/mnras/stw145}
  {\path{doi:10.1093/mnras/stw145}}.

\bibitem{Pawlowski:2015qta}
M.~S. Pawlowski, B.~Famaey, D.~Merritt, P.~Kroupa, {On the persistence of two
  small-scale problems in $\Lambda$CDM}, Astrophys. J. 815~(1) (2015) 19.
\newblock \href {http://arxiv.org/abs/1510.08060} {\path{arXiv:1510.08060}},
  \href {http://dx.doi.org/10.1088/0004-637X/815/1/19}
  {\path{doi:10.1088/0004-637X/815/1/19}}.

\bibitem{Mastache:2019bxu}
J.~Mastache, A.~de~la Macorra, {Bound Dark Matter (BDM) towards solving the
  Small Scale Structure Problem}, JCAP 2003~(03) (2020) 025.
\newblock \href {http://arxiv.org/abs/1909.00488} {\path{arXiv:1909.00488}},
  \href {http://dx.doi.org/10.1088/1475-7516/2020/03/025}
  {\path{doi:10.1088/1475-7516/2020/03/025}}.

\bibitem{Bode:2000gq}
P.~Bode, J.~P. Ostriker, N.~Turok, {Halo formation in warm dark matter models},
  Astrophys. J. 556 (2001) 93--107.
\newblock \href {http://arxiv.org/abs/astro-ph/0010389}
  {\path{arXiv:astro-ph/0010389}}, \href {http://dx.doi.org/10.1086/321541}
  {\path{doi:10.1086/321541}}.

\bibitem{Colin:2000dn}
P.~Colin, V.~Avila-Reese, O.~Valenzuela, {Substructure and halo density
  profiles in a warm dark matter cosmology}, Astrophys. J. 542 (2000) 622--630.
\newblock \href {http://arxiv.org/abs/astro-ph/0004115}
  {\path{arXiv:astro-ph/0004115}}, \href {http://dx.doi.org/10.1086/317057}
  {\path{doi:10.1086/317057}}.

\bibitem{Hansen:2001zv}
S.~H. Hansen, J.~Lesgourgues, S.~Pastor, J.~Silk, {Constraining the window on
  sterile neutrinos as warm dark matter}, Mon. Not. Roy. Astron. Soc. 333
  (2002) 544--546.
\newblock \href {http://arxiv.org/abs/astro-ph/0106108}
  {\path{arXiv:astro-ph/0106108}}, \href
  {http://dx.doi.org/10.1046/j.1365-8711.2002.05410.x}
  {\path{doi:10.1046/j.1365-8711.2002.05410.x}}.

\bibitem{Viel:2005qj}
M.~Viel, J.~Lesgourgues, M.~G. Haehnelt, S.~Matarrese, A.~Riotto, {Constraining
  warm dark matter candidates including sterile neutrinos and light gravitinos
  with WMAP and the Lyman-alpha forest}, Phys. Rev. D71 (2005) 063534.
\newblock \href {http://arxiv.org/abs/astro-ph/0501562}
  {\path{arXiv:astro-ph/0501562}}, \href
  {http://dx.doi.org/10.1103/PhysRevD.71.063534}
  {\path{doi:10.1103/PhysRevD.71.063534}}.

\bibitem{Dodelson:1993je}
S.~Dodelson, L.~M. Widrow, {Sterile-neutrinos as dark matter}, Phys. Rev. Lett.
  72 (1994) 17--20.
\newblock \href {http://arxiv.org/abs/hep-ph/9303287}
  {\path{arXiv:hep-ph/9303287}}, \href
  {http://dx.doi.org/10.1103/PhysRevLett.72.17}
  {\path{doi:10.1103/PhysRevLett.72.17}}.

\bibitem{Dolgov:2000ew}
A.~D. Dolgov, S.~H. Hansen, {Massive sterile neutrinos as warm dark matter},
  Astropart. Phys. 16 (2002) 339--344.
\newblock \href {http://arxiv.org/abs/hep-ph/0009083}
  {\path{arXiv:hep-ph/0009083}}, \href
  {http://dx.doi.org/10.1016/S0927-6505(01)00115-3}
  {\path{doi:10.1016/S0927-6505(01)00115-3}}.

\bibitem{Asaka:2006nq}
T.~Asaka, M.~Laine, M.~Shaposhnikov, {Lightest sterile neutrino abundance
  within the nuMSM}, JHEP 01 (2007) 091, [Erratum: JHEP02,028(2015)].
\newblock \href {http://arxiv.org/abs/hep-ph/0612182}
  {\path{arXiv:hep-ph/0612182}}, \href
  {http://dx.doi.org/10.1088/1126-6708/2007/01/091, 10.1007/JHEP02(2015)028}
  {\path{doi:10.1088/1126-6708/2007/01/091, 10.1007/JHEP02(2015)028}}.

\bibitem{Shi:1998km}
X.-D. Shi, G.~M. Fuller, {A New dark matter candidate: Nonthermal sterile
  neutrinos}, Phys. Rev. Lett. 82 (1999) 2832--2835.
\newblock \href {http://arxiv.org/abs/astro-ph/9810076}
  {\path{arXiv:astro-ph/9810076}}, \href
  {http://dx.doi.org/10.1103/PhysRevLett.82.2832}
  {\path{doi:10.1103/PhysRevLett.82.2832}}.

\bibitem{Abazajian:2001nj}
K.~Abazajian, G.~M. Fuller, M.~Patel, {Sterile neutrino hot, warm, and cold
  dark matter}, Phys. Rev. D64 (2001) 023501.
\newblock \href {http://arxiv.org/abs/astro-ph/0101524}
  {\path{arXiv:astro-ph/0101524}}, \href
  {http://dx.doi.org/10.1103/PhysRevD.64.023501}
  {\path{doi:10.1103/PhysRevD.64.023501}}.

\bibitem{Kusenko:2006rh}
A.~Kusenko, {Sterile neutrinos, dark matter, and the pulsar velocities in
  models with a Higgs singlet}, Phys. Rev. Lett. 97 (2006) 241301.
\newblock \href {http://arxiv.org/abs/hep-ph/0609081}
  {\path{arXiv:hep-ph/0609081}}, \href
  {http://dx.doi.org/10.1103/PhysRevLett.97.241301}
  {\path{doi:10.1103/PhysRevLett.97.241301}}.

\bibitem{Petraki:2007gq}
K.~Petraki, A.~Kusenko, {Dark-matter sterile neutrinos in models with a gauge
  singlet in the Higgs sector}, Phys. Rev. D77 (2008) 065014.
\newblock \href {http://arxiv.org/abs/0711.4646} {\path{arXiv:0711.4646}},
  \href {http://dx.doi.org/10.1103/PhysRevD.77.065014}
  {\path{doi:10.1103/PhysRevD.77.065014}}.

\bibitem{Merle:2015oja}
A.~Merle, M.~Totzauer, {keV Sterile Neutrino Dark Matter from Singlet Scalar
  Decays: Basic Concepts and Subtle Features}, JCAP 1506 (2015) 011.
\newblock \href {http://arxiv.org/abs/1502.01011} {\path{arXiv:1502.01011}},
  \href {http://dx.doi.org/10.1088/1475-7516/2015/06/011}
  {\path{doi:10.1088/1475-7516/2015/06/011}}.

\bibitem{Konig:2016dzg}
J.~König, A.~Merle, M.~Totzauer, {keV Sterile Neutrino Dark Matter from Singlet
  Scalar Decays: The Most General Case}, JCAP 1611~(11) (2016) 038.
\newblock \href {http://arxiv.org/abs/1609.01289} {\path{arXiv:1609.01289}},
  \href {http://dx.doi.org/10.1088/1475-7516/2016/11/038}
  {\path{doi:10.1088/1475-7516/2016/11/038}}.

\bibitem{Muller:2004yb}
C.~M. Muller, {Cosmological bounds on the equation of state of dark matter},
  Phys. Rev. D71 (2005) 047302.
\newblock \href {http://arxiv.org/abs/astro-ph/0410621}
  {\path{arXiv:astro-ph/0410621}}, \href
  {http://dx.doi.org/10.1103/PhysRevD.71.047302}
  {\path{doi:10.1103/PhysRevD.71.047302}}.

\bibitem{Faber:2005xc}
T.~Faber, M.~Visser, {Combining rotation curves and gravitational lensing: How
  to measure the equation of state of dark matter in the galactic halo}, Mon.
  Not. Roy. Astron. Soc. 372 (2006) 136--142.
\newblock \href {http://arxiv.org/abs/astro-ph/0512213}
  {\path{arXiv:astro-ph/0512213}}, \href
  {http://dx.doi.org/10.1111/j.1365-2966.2006.10845.x}
  {\path{doi:10.1111/j.1365-2966.2006.10845.x}}.

\bibitem{Serra:2011jh}
A.~L. Serra, M.~J. d. L.~D. Romero, {Measuring the dark matter equation of
  state}, Mon. Not. Roy. Astron. Soc. 415 (2011) 74.
\newblock \href {http://arxiv.org/abs/1103.5465} {\path{arXiv:1103.5465}},
  \href {http://dx.doi.org/10.1111/j.1745-3933.2011.01082.x}
  {\path{doi:10.1111/j.1745-3933.2011.01082.x}}.

\bibitem{Hu:1998kj}
W.~Hu, {Structure formation with generalized dark matter}, Astrophys. J. 506
  (1998) 485--494.
\newblock \href {http://arxiv.org/abs/astro-ph/9801234}
  {\path{arXiv:astro-ph/9801234}}, \href {http://dx.doi.org/10.1086/306274}
  {\path{doi:10.1086/306274}}.

\bibitem{Kopp:2016mhm}
M.~Kopp, C.~Skordis, D.~B. Thomas, {Extensive investigation of the generalized
  dark matter model}, Phys. Rev. D94~(4) (2016) 043512.
\newblock \href {http://arxiv.org/abs/1605.00649} {\path{arXiv:1605.00649}},
  \href {http://dx.doi.org/10.1103/PhysRevD.94.043512}
  {\path{doi:10.1103/PhysRevD.94.043512}}.

\bibitem{Thomas:2016iav}
D.~B. Thomas, M.~Kopp, C.~Skordis, {Constraining the Properties of Dark Matter
  with Observations of the Cosmic Microwave Background}, Astrophys. J. 830~(2)
  (2016) 155.
\newblock \href {http://arxiv.org/abs/1601.05097} {\path{arXiv:1601.05097}},
  \href {http://dx.doi.org/10.3847/0004-637X/830/2/155}
  {\path{doi:10.3847/0004-637X/830/2/155}}.

\bibitem{Kopp:2018zxp}
M.~Kopp, C.~Skordis, D.~B. Thomas, S.~Ili?, {Dark Matter Equation of State
  through Cosmic History}, Phys. Rev. Lett. 120~(22) (2018) 221102.
\newblock \href {http://arxiv.org/abs/1802.09541} {\path{arXiv:1802.09541}},
  \href {http://dx.doi.org/10.1103/PhysRevLett.120.221102}
  {\path{doi:10.1103/PhysRevLett.120.221102}}.

\bibitem{Menci:2016eui}
N.~Menci, A.~Grazian, M.~Castellano, N.~G. Sanchez, {A Stringent Limit on the
  Warm Dark Matter Particle Masses from the Abundance of z=6 Galaxies in the
  Hubble Frontier Fields}, Astrophys. J. 825~(1) (2016) L1.
\newblock \href {http://arxiv.org/abs/1606.02530} {\path{arXiv:1606.02530}},
  \href {http://dx.doi.org/10.3847/2041-8205/825/1/L1}
  {\path{doi:10.3847/2041-8205/825/1/L1}}.

\bibitem{Corasaniti:2016epp}
P.~S. Corasaniti, S.~Agarwal, D.~J.~E. Marsh, S.~Das, {Constraints on dark
  matter scenarios from measurements of the galaxy luminosity function at high
  redshifts}, Phys. Rev. D95~(8) (2017) 083512.
\newblock \href {http://arxiv.org/abs/1611.05892} {\path{arXiv:1611.05892}},
  \href {http://dx.doi.org/10.1103/PhysRevD.95.083512}
  {\path{doi:10.1103/PhysRevD.95.083512}}.

\bibitem{Pacucci:2013jfa}
F.~Pacucci, A.~Mesinger, Z.~Haiman, {Focusing on Warm Dark Matter with Lensed
  High-redshift Galaxies}, Mon. Not. Roy. Astron. Soc. 435 (2013) L53.
\newblock \href {http://arxiv.org/abs/1306.0009} {\path{arXiv:1306.0009}},
  \href {http://dx.doi.org/10.1093/mnrasl/slt093}
  {\path{doi:10.1093/mnrasl/slt093}}.

\bibitem{Viel:2013apy}
M.~Viel, G.~D. Becker, J.~S. Bolton, M.~G. Haehnelt, {Warm dark matter as a
  solution to the small scale crisis: New constraints from high redshift
  Lyman-α forest data}, Phys. Rev. D88 (2013) 043502.
\newblock \href {http://arxiv.org/abs/1306.2314} {\path{arXiv:1306.2314}},
  \href {http://dx.doi.org/10.1103/PhysRevD.88.043502}
  {\path{doi:10.1103/PhysRevD.88.043502}}.

\bibitem{Hui:2016ltb}
L.~Hui, J.~P. Ostriker, S.~Tremaine, E.~Witten, {Ultralight scalars as
  cosmological dark matter}, Phys. Rev. D95~(4) (2017) 043541.
\newblock \href {http://arxiv.org/abs/1610.08297} {\path{arXiv:1610.08297}},
  \href {http://dx.doi.org/10.1103/PhysRevD.95.043541}
  {\path{doi:10.1103/PhysRevD.95.043541}}.

\bibitem{Ma:1995ey}
C.-P. Ma, E.~Bertschinger, {Cosmological perturbation theory in the synchronous
  and conformal Newtonian gauges}, Astrophys. J. 455 (1995) 7--25.
\newblock \href {http://arxiv.org/abs/astro-ph/9506072}
  {\path{arXiv:astro-ph/9506072}}, \href {http://dx.doi.org/10.1086/176550}
  {\path{doi:10.1086/176550}}.

\bibitem{Lesgourgues:2011rh}
J.~Lesgourgues, T.~Tram, {The Cosmic Linear Anisotropy Solving System (CLASS)
  IV: efficient implementation of non-cold relics}, JCAP 1109 (2011) 032.
\newblock \href {http://arxiv.org/abs/1104.2935} {\path{arXiv:1104.2935}},
  \href {http://dx.doi.org/10.1088/1475-7516/2011/09/032}
  {\path{doi:10.1088/1475-7516/2011/09/032}}.

\bibitem{Blas:2011rf}
D.~Blas, J.~Lesgourgues, T.~Tram, {The Cosmic Linear Anisotropy Solving System
  (CLASS) II: Approximation schemes}, JCAP 1107 (2011) 034.
\newblock \href {http://arxiv.org/abs/1104.2933} {\path{arXiv:1104.2933}},
  \href {http://dx.doi.org/10.1088/1475-7516/2011/07/034}
  {\path{doi:10.1088/1475-7516/2011/07/034}}.

\bibitem{Parkinson:2012vd}
D.~Parkinson, et~al., {The WiggleZ Dark Energy Survey: Final data release and
  cosmological results}, Phys. Rev. D 86 (2012) 103518.
\newblock \href {http://arxiv.org/abs/1210.2130} {\path{arXiv:1210.2130}},
  \href {http://dx.doi.org/10.1103/PhysRevD.86.103518}
  {\path{doi:10.1103/PhysRevD.86.103518}}.

\bibitem{MacorraVirial:2019gui}
A.~de~la Macorra, J.~Mastache, {The virial mode $k_v$ approach to Structure
  Formation with Warm Dark Matter}\href {http://arxiv.org/abs/2009.05745}
  {\path{arXiv:2009.05745}}.

\bibitem{Press:1973iz}
W.~H. Press, P.~Schechter, {Formation of galaxies and clusters of galaxies by
  selfsimilar gravitational condensation}, Astrophys. J. 187 (1974) 425--438.
\newblock \href {http://dx.doi.org/10.1086/152650} {\path{doi:10.1086/152650}}.

\bibitem{Benson:2012su}
A.~J. Benson, A.~Farahi, S.~Cole, L.~A. Moustakas, A.~Jenkins, M.~Lovell,
  R.~Kennedy, J.~Helly, C.~Frenk, {Dark Matter Halo Merger Histories Beyond
  Cold Dark Matter: I - Methods and Application to Warm Dark Matter}, Mon. Not.
  Roy. Astron. Soc. 428 (2013) 1774.
\newblock \href {http://arxiv.org/abs/1209.3018} {\path{arXiv:1209.3018}},
  \href {http://dx.doi.org/10.1093/mnras/sts159}
  {\path{doi:10.1093/mnras/sts159}}.

\bibitem{Bond:1990iw}
J.~R. Bond, S.~Cole, G.~Efstathiou, N.~Kaiser, {Excursion set mass functions
  for hierarchical Gaussian fluctuations}, Astrophys. J. 379 (1991) 440.
\newblock \href {http://dx.doi.org/10.1086/170520} {\path{doi:10.1086/170520}}.

\bibitem{Markovic:2010te}
K.~Markovic, S.~Bridle, A.~Slosar, J.~Weller, {Constraining warm dark matter
  with cosmic shear power spectra}, JCAP 1101 (2011) 022.
\newblock \href {http://arxiv.org/abs/1009.0218} {\path{arXiv:1009.0218}},
  \href {http://dx.doi.org/10.1088/1475-7516/2011/01/022}
  {\path{doi:10.1088/1475-7516/2011/01/022}}.

\bibitem{vandenBosch:2003nk}
F.~C. van~den Bosch, H.~J. Mo, X.~Yang, {Towards cosmological concordance on
  galactic scales}, Mon. Not. Roy. Astron. Soc. 345 (2003) 923.
\newblock \href {http://arxiv.org/abs/astro-ph/0301104}
  {\path{arXiv:astro-ph/0301104}}, \href
  {http://dx.doi.org/10.1046/j.1365-8711.2003.07012.x}
  {\path{doi:10.1046/j.1365-8711.2003.07012.x}}.

\bibitem{Murgia:2017lwo}
R.~Murgia, A.~Merle, M.~Viel, M.~Totzauer, A.~Schneider, {"Non-cold" dark
  matter at small scales: a general approach}, JCAP 1711 (2017) 046.
\newblock \href {http://arxiv.org/abs/1704.07838} {\path{arXiv:1704.07838}},
  \href {http://dx.doi.org/10.1088/1475-7516/2017/11/046}
  {\path{doi:10.1088/1475-7516/2017/11/046}}.

\end{thebibliography}

\appendix
\section{Energy density at $a \simeq a_{nr}$}\label{app:rho}

We can obtain the energy density as function of the temperature $T$, if we weight each state by the $E(p) = \left( m^2 + p^2  \right)^{1/2} \approx (m+p)/\sqrt{2}$ which is upto the second term in a Taylor expansion when $p \approx m$, this is, when $a \approx a_{nr}$. Therefore, the energy density is given by
\begin{eqnarray}
    \rho &=& \frac{g}{(2 \pi)^{3}} \int d^{3} p f(p) E(p)  \\    
    \rho_{\rm bos}(m,T) &\approx& \frac{\sqrt{2}\pi^2\,g T^4}{15}\left( 1 - \frac{30\sqrt{2}\xi(3)}{\pi^4} \fr{m}{T} + \frac{5}{4\pi^2}\frac{m^2}{T^2} \right)    \label{eq:rho_boson}    \\    
    \rho_{\rm fer}(m, T) &\approx&  \frac{7}{8} \frac{\sqrt{2}\pi^2 g T^4}{15}  \left( 1 - \frac{8}{7}\frac{45}{2}\frac{\sqrt{2}\xi(3)}{\pi^4}\frac{m}{T} + \frac{5}{7\pi^2}\frac{m^2}{T^2} \right)      \label{eq:rho_fermion1}
\end{eqnarray}
where $\rho_{\rm bos},\;\rho_{\rm fer}$ stand for bosons and fermions, respectively. We can see, that the Eq.\eqref{eq:rho_boson} and \eqref{eq:rho_fermion1} have the classic relation $\rho_{\rm fer} = \frac{7}{8}\rho_{\rm bos}$ for the first term in the parenthesis, although the proportionality value between the other terms in both equations have different values. Let us define the relation between the temperature and the mass as $T = \sigma m$, which is valid for the massive non-relativistic limit, in which case have $T = \omega m$, where $\omega$ came from the equation of state. Then, for fermions Eq.\eqref{eq:rho_fermion1} can be rewritten as,
\begin{equation}
     \rho_{\rm fer}(T) = \frac{7}{8} \frac{g T^4}{\sqrt{2}\pi^2}\left( \frac{2\pi^4}{15} - \frac{24\sqrt{2}}{7\sigma}\xi(3) + \frac{2 \pi^2}{21\sigma^2} \right)
\end{equation} \label{eq:rho_fermion3}
where $\xi$ is the Riemann zeta-function. Everything in the parenthesis is constant, for instance, for $\sigma = 1/3$, the value of the parenthesis in Eqs.\eqref{eq:rho_boson} y\eqref{eq:rho_fermion1} is  7.39 (3.46) for boson (fermion).

\section{Transfer function and the Matter Power Spectrum}\la{app.alpha}
The Matter Power Spectrum for WDM can be computed from a parametrization of the transfer function given in  Eq.\eqref{eq:transfer_viel} \cite{Bode:2000gq, Colin:2000dn, Hansen:2001zv,Viel:2005qj}. Inspired in this parametrization we propose the virial transfer function with a clear conection to  large scale structure formation \cite{MacorraVirial:2019gui}, Eq.\eqref{eq:transf_virial}, using all the physical concepts introduce in We also found that the following parametrization may also be a good candidate for the transfer function,
\begin{equation}\label{eq:second_transfer}
	T =   \left(1 + \left(\frac{k}{k_{fs}} \right)^{2\beta}\right)^{-3/\beta}
\end{equation}
\begin{equation}
	\beta = 1.12 \pm 0.04
\end{equation}
Using the value for $a_{nr}$ we compute $k_{fs}$ by determining first $\lambda_{fs}$ using Eqs.\eqref{eq:lambda_fs}. Notice that the standard error for $\beta$ is small. The value of $2\beta \simeq 2.23$ is actually close to previous works \cite{Murgia:2017lwo} that obtained $2\nu \simeq 2.24$, however we found a difference between the values $-3/\beta = -2.68$ and $-5/\nu = -4.46$. We also want to highlight that the $\alpha$ parameter in Eq.\eqref{eq:transf_virial} lacks the connection to the physics of the WDM particle, while in Eq.\eqref{eq:second_transfer} all the physics is encoded in $k_{fs}$ and the free-streaming scale concept. Parametrization in Eq.\eqref{eq:second_transfer} may be explorer in a future work.

\end{document}